# Photonic zitterbewegung and its interpretation*

Zhi-Yong Wang[†], Cai-Dong Xiong, Qi Qiu

*School of Optoelectronic Information, University of Electronic Science and Technology of China, Chengdu 610054, CHINA*

In term of the volume-integrated Poynting vector, we present a quantum field-theory investigation on the zitterbewegung (ZB) of photons, and show that this ZB occurs only in the presence of virtual longitudinal and scalar photons. To present a heuristic explanation for such ZB, by assuming that the space time is sufficiently close to the flat Minkowski space, we show that the gravitational interaction can result in the ZB of photons.

## 1. Introduction

Zitterbewegung (ZB, English: "trembling motion", from German) is a theoretical rapid motion of elementary particles. The existence of such motion was first proposed by E. Schrödinger in 1930 as a result of his analysis of the wave packet solutions of the Dirac equation for relativistic electrons in free space, in which an interference between positive and negative energy states produces what appears to be a fluctuation of the position of an electron around the median. Historically, the investigations on the ZB of relativistic particles are mainly focused on the Dirac electron, and have been developed just at the level of quantum mechanics. Some investigations on the ZB of photons are based on the concept of position operator (e.g., Ref. [1]), unfortunately, the photonic position operator is an ill-defined concept. Recently, an old interest in ZB has been rekindled by the investigations on spintronics, graphene, and superconducting systems, etc. [2-10], which inspires us to present a quantum field-theory investigation on the ZB of the Dirac particles [11-12].

---
* Supported by the Fundamental Research Funds for the Central Universities (Grant No: ZYGX2010X013).
[†]Corresponding author. E-mail: zywang@uestc.edu.cn

As for a free Dirac electron (with the mass of *m*) in a vacuum, the ZB frequency is of the order of $mc^2/\hbar$ ($\approx 10^{20}$ Hz), and the spatial extension of the ZB motion comparable to the Compton wavelength $\hbar/mc$ ($\approx 10^{-12}$ m). Thus, it is impossible to observe this effect with the currently available experimental means. However, the presence of ZB is a general feature of spinor systems with linear dispersion relations. Trapped ions as well as condensed matter systems, including graphene and semiconductor quantum wires, have been proposed as candidate systems for observing ZB [2-16]. For example, the ZB of electrons in narrow gap semiconductors having much more advantageous characteristics [14]: the frequency $\omega_Z = \varepsilon_g/\hbar$ ($\approx 10^{15}$ Hz) and the amplitude $\lambda_Z = \hbar/m_0^* u$ ($\approx 10^{-9}$ m), where $\varepsilon_g$ is the energy gap, $m_0^*$ is the effective mass, and $u \approx 10^6$ m/s (plays the role of velocity of light in vacuum) is the maximum electron velocity in the two-band energy spectrum. One can expect that the ZB of charge carriers in these materials can be observed experimentally. In fact, it was recently claimed that ZB has now been measured [8]. Moreover, an observation of an acoustic analogue of ZB was reported recently in sonic crystals [17].

On the other hand, in Ref. [18], the author claimed that the ZB of photons can appear near the Dirac point in two-dimensional photonic crystal, and it can be observed by measuring the time dependence of the transmission coefficient through photonic crystal slabs. However, in contrary to the theory of the Dirac electron's ZB, a rigorous theory for photons' ZB is still absent (the existing theory of photons' ZB is problematical: 1) it is based on an ill-defined concept, i.e., photonic position operator; 2) it is based on quantum mechanics, but the quantum theory of photons is directly a quantum field theory). As a result, the validity of conclusions in Ref. [18] is easy to be oppugned.

In this paper, applying the volume-integrated Poynting vector, we present a quantum field-theory investigation on the ZB of photons, and present a heuristic explanation for such ZB. In the following the natural units of $\hbar = c = 1$ are applied, and the four-dimensional (4D) flat space-time metric tensor is taken as $\eta_{\mu\nu} = \text{diag}(1,-1,-1,-1)$ ($\mu,\nu = 0,1,2,3$). Repeated indices must be summed according to the Einstein rule.

## 2. Photonic ZB

In general, the ZB of particles can be studied in terms of the corresponding free fields in flat spacetime, while, to interpret how the ZB behaviors occur, one has to resort to the interactions between quantum fields. In a flat spacetime and outside a source, let $A^\mu(x) = (A^0, \boldsymbol{A})$ be a 4D electromagnetic potential, in terms of which the electromagnetic field intensities can be expressed as

$$\boldsymbol{E} = -\nabla A^0 - \partial_t \boldsymbol{A}, \quad \boldsymbol{B} = \nabla \times \boldsymbol{A}. \tag{1}$$

In the Lorentz gauge condition, one can expand $A^\mu(x)$ as

$$A^\mu(x) = \sum_{\boldsymbol{k}} \sum_{s=0}^{3} \frac{1}{\sqrt{2\omega V}} e^\mu(\boldsymbol{k}, s)[\hat{b}(\boldsymbol{k}, s)\exp(-\mathrm{i}k_\mu x^\mu) + \hat{b}^\dagger(\boldsymbol{k}, s)\exp(\mathrm{i}k_\mu x^\mu)], \tag{2}$$

where $V = \int \mathrm{d}^3 x$ represents the normalization volume (the box normalization is adopted), $k_\mu x^\mu = \omega t - \boldsymbol{k} \cdot \boldsymbol{x}$, $k^\mu = (\omega, \boldsymbol{k})$ (with $\omega = |\boldsymbol{k}|$) represents the 4D wavenumber vector (also the 4D momentum of single photons, $\hbar = c = 1$), the expansion coefficients $\hat{b}(\boldsymbol{k}, s)$ and $\hat{b}^\dagger(\boldsymbol{k}, s)$ ($\hat{b}^\dagger$ denotes the hermitian conjugate of $\hat{b}$, and so on) represent the photonic annihilation and creation operators, respectively. Moreover, the unit vectors $e^\mu(\boldsymbol{k}, s)$'s are four 4D polarization vectors with the four indices of $s=0,1,2,3$ corresponding to four kinds of photons, respectively. Let $\{\boldsymbol{\varepsilon}(\boldsymbol{k}, 1), \boldsymbol{\varepsilon}(\boldsymbol{k}, -1), \boldsymbol{\varepsilon}(\boldsymbol{k}, 0)\}$ denote a 3D orthonormal basis, which is formed by the following three unit vectors

$$\begin{cases} \boldsymbol{\varepsilon}(\boldsymbol{k}, 1) = \boldsymbol{\varepsilon}^*(\boldsymbol{k}, -1) = \dfrac{1}{\sqrt{2}|\boldsymbol{k}|}\left(\dfrac{k_1 k_3 - \mathrm{i}k_2|\boldsymbol{k}|}{k_1 - \mathrm{i}k_2}, \dfrac{k_2 k_3 + \mathrm{i}k_1|\boldsymbol{k}|}{k_1 - \mathrm{i}k_2}, -(k_1 + \mathrm{i}k_2)\right) \\ \boldsymbol{\varepsilon}(\boldsymbol{k}, 0) = \dfrac{\boldsymbol{k}}{|\boldsymbol{k}|} = \dfrac{1}{|\boldsymbol{k}|}(k_1, k_2, k_3) \end{cases}, \tag{3}$$

where $\boldsymbol{\varepsilon}^*(\boldsymbol{k}, -1)$ denotes the complex conjugate of $\boldsymbol{\varepsilon}(\boldsymbol{k}, -1)$ (and so on). As an example, let $\boldsymbol{k} = (0, 0, k_3)$ with $k_3 > 0$, one has $\boldsymbol{\varepsilon}(\boldsymbol{k}, \pm 1) = \sqrt{1/2}(1, \pm \mathrm{i}, 0)$ and $\boldsymbol{\varepsilon}(\boldsymbol{k}, 0) = (0, 0, 1)$. In fact, $\boldsymbol{\varepsilon}(\boldsymbol{k}, \pm 1)$ and $\boldsymbol{\varepsilon}(\boldsymbol{k}, 0)$ stand for the three polarization vectors of the electromagnetic field, where $\boldsymbol{\varepsilon}(\boldsymbol{k}, 0)$ (parallel to $\boldsymbol{k}$) denotes the longitudinal polarization vector, while $\boldsymbol{\varepsilon}(\boldsymbol{k}, \pm 1)$ (perpendicular to $\boldsymbol{k}$) correspond to the right- and left-hand circular polarization vectors, respectively. Therefore, they are actually the spinor representation of three linear polarization vectors. Using Eq. (3) with $\omega = |\boldsymbol{k}|$, the four 4D polarization vectors $e^\mu(\boldsymbol{k},s)$ ($s = 0,1,2,3$) in Eq. (2) can be chosen as

$$\begin{cases} e^\mu(\boldsymbol{k},0) = (1,0,0,0), & e^\mu(\boldsymbol{k},1) = (0, \varepsilon(\boldsymbol{k},1)) \\ e^\mu(\boldsymbol{k},2) = (0, \varepsilon(\boldsymbol{k},-1)), & e^\mu(\boldsymbol{k},3) = (0, \varepsilon(\boldsymbol{k},0)) \end{cases}. \quad (4)$$

Applying Eqs. (1)-(4). $\boldsymbol{k} = |\boldsymbol{k}|\varepsilon(\boldsymbol{k},0) = \omega\varepsilon(\boldsymbol{k},0)$, and $\lambda\varepsilon(\boldsymbol{k},\lambda) = 0$ for $\lambda = 0$, one can show that the electromagnetic field intensities are:

$$\begin{cases} \boldsymbol{E} = \sum_{\boldsymbol{k},\lambda} \sqrt{\frac{\omega}{V}} [\frac{1}{\sqrt{1+\lambda^2}} \varepsilon(\boldsymbol{k},\lambda)][\hat{a}(\boldsymbol{k},\lambda)\exp(-ik_\mu x^\mu) + \hat{a}^\dagger(\boldsymbol{k},\lambda)\exp(ik_\mu x^\mu)] \\ \boldsymbol{B} = \sum_{\boldsymbol{k},\lambda} \sqrt{\frac{\omega}{V}} [\frac{-i}{\sqrt{1+\lambda^2}} \lambda\varepsilon(\boldsymbol{k},\lambda)][\hat{a}(\boldsymbol{k},\lambda)\exp(-ik_\mu x^\mu) + \hat{a}^\dagger(\boldsymbol{k},\lambda)\exp(ik_\mu x^\mu)] \end{cases}, \quad (5)$$

where $\lambda = \pm 1, 0$, and

$$\begin{cases} \hat{a}(\boldsymbol{k},1) = i\hat{b}(\boldsymbol{k},1) \\ \hat{a}(\boldsymbol{k},-1) = i\hat{b}(\boldsymbol{k},2) \\ \hat{a}(\boldsymbol{k},0) = i[\hat{b}(\boldsymbol{k},3) - \hat{b}(\boldsymbol{k},0)]/\sqrt{2} \end{cases}. \quad (6)$$

Therefore, when the electromagnetic field is described by the 4D electromagnetic potential, there involves four polarization vectors together describing four kinds of photons; while described by the electromagnetic field intensities, there only involves three polarization vectors labeled by the parameters $\lambda = \pm 1, 0$, and Eq. (6) implies that the $\lambda = 0$ solution describes the admixture of the longitudinal (s=3) and scalar (s=0) photons, and the $\lambda = \pm 1$ solutions describe the two transverse photons (s=1, 2). According to QED, only those state vectors (say, $|\Phi\rangle$) are admitted for which the expectation value of the Lorentz gauge condition is satisfied: $\langle\Phi|\partial^\mu A_\mu|\Phi\rangle = 0$. Let $A_\mu(x) = A_\mu^{(+)}(x) + A_\mu^{(-)}(x)$, where $A_\mu^{(+)}$ and $A_\mu^{(-)}$ are the positive and negative frequency components of $A_\mu(x)$, respectively, applying the stronger condition of $\partial^\mu A_\mu^{(+)}|\Phi\rangle = 0$, consider that $k_\mu e^\mu(\boldsymbol{k},1) = k_\mu e^\mu(\boldsymbol{k},2) = 0$ and $k_\mu e^\mu(\boldsymbol{k},0) = -k_\mu e^\mu(\boldsymbol{k},3)$, one can obtain

$$\hat{a}(\boldsymbol{k},0)|\Phi\rangle = \mathrm{i}\frac{1}{\sqrt{2}}[\hat{b}(\boldsymbol{k},3)-\hat{b}(\boldsymbol{k},0)]|\Phi\rangle = 0. \tag{7-1}$$

Similarly, using $\langle\Phi|\partial^\mu A_\mu^{(-)}=0$ one has

$$\langle\Phi|\hat{a}^\dagger(\boldsymbol{k},0) = \langle\Phi|\mathrm{i}\frac{1}{\sqrt{2}}[\hat{b}^\dagger(\boldsymbol{k},3)-\hat{b}^\dagger(\boldsymbol{k},0)] = 0. \tag{7-2}$$

According to QED, there are the commutation relations:

$$[\hat{b}(\boldsymbol{k},s),\hat{b}^\dagger(\boldsymbol{k}',s')]=-\eta_{ss'}\delta_{\boldsymbol{kk}'}, \tag{8}$$

with the others vanishing. Applying Eqs. (5) and (7), one can obtain

$$[\hat{a}(k,\lambda),\hat{a}^\dagger(k',\lambda')]=\delta_{\boldsymbol{kk}'}\delta_{\lambda\lambda'},\ \lambda,\lambda'=\pm 1, \tag{9}$$

with other commutators vanishing. In particular, one has $[\hat{a}(k,0),\hat{a}^\dagger(k',0)]=0$. The ZB of photons can be studied via the momentum of the electromagnetic field given by $\boldsymbol{J}=\int(\boldsymbol{E}\times\boldsymbol{B})\mathrm{d}^3x$. Using Eqs. (5) and (9) one has

$$\begin{aligned}\boldsymbol{J} &= \int(\boldsymbol{E}\times\boldsymbol{B})\mathrm{d}^3x \\ &= \sum_{\boldsymbol{k},\lambda=\pm 1}\boldsymbol{k}\hat{a}^\dagger(k,\lambda)\hat{a}(k,\lambda) - \sum_{\boldsymbol{k},\lambda=\pm 1}\frac{\omega}{\sqrt{2}}\boldsymbol{\varepsilon}(\boldsymbol{k},\lambda)[\hat{a}^\dagger(k,0)\hat{a}(k,\lambda)+\mathrm{h.c.}] \\ &\quad -\sum_{\boldsymbol{k},\lambda=\pm 1}\frac{\lambda\omega}{2\sqrt{2}}\boldsymbol{\varepsilon}(\boldsymbol{k},-\lambda)[\hat{a}(k,\lambda)\hat{a}(-k,0)\exp(-2\mathrm{i}\omega t)+\mathrm{h.c.}] \\ &\quad +\sum_{\boldsymbol{k},\lambda=\pm 1}\frac{\lambda\omega}{2\sqrt{2}}\boldsymbol{\varepsilon}(-\boldsymbol{k},\lambda)[\hat{a}(k,0)\hat{a}(-k,\lambda)\exp(-2\mathrm{i}\omega t)+\mathrm{h.c.}]\end{aligned} \tag{10}$$

where $\hat{a}(\pm k,0)\equiv\hat{a}(k_0,\pm\boldsymbol{k},0)$ (and so on). The first term on the right-hand side of Eq. (10), as the classic term, represents the total momentum of transverse photons (i.e., radiation photons), while the others are usually left out because of Eq. (7). On the other hand, the third and fourth terms on the right-hand side of Eq. (10) are perpendicular to $\boldsymbol{k}$ and stand for the transverse ZB motions superposing the classic momentum, they contain $\hat{a}(\pm k,0)$ and $\hat{a}^\dagger(\pm k,0)$ and then correspond to the admixture of longitudinal and scalar photons (see Eq. (6)). Eq. (7) implies that the contributions of longitudinal and scalar photons cancel each other. However, the longitudinal and scalar photons can exist in the form of virtual photons (for example, the Coulomb interaction arises from the combined exchange of *virtual* longitudinal and scalar photons [19]). Therefore, Eq. (10) shows that the ZB of photons occurs only in the presence of virtual longitudinal and scalar photons.

## 3. A heuristic interpretation for photonic ZB

In Refs. [20, 21], an interpretation for the Dirac electron's ZB is presented in quantum-mechanical language, which can be restated in the language of quantum field theory [12], i.e., the ZB for a particle arises from the influence of virtual particle-antiparticle pairs around the particle. In a similar way one can present an interpretation for photons' ZB. However, photons do not carry the electric charge (only carry the gravitational charge, i.e., photons' 4D momentum), to induce the ZB of photons, the electromagnetic vacuum should be replaced by the gravitational vacuum, where the gravitational field is related to a photon in the similar way that the electromagnetic field is related to an electron. In fact, the contributions from the ZB terms of Eq. (10) do not vanish provided that Eq. (7) does not hold true. Then, let us discuss the ZB of photons in a gravitational field (i.e., in a curved spacetime). For simplicity, we will just provide a heuristic explanation.

Eq. (2) can be regarded as an expansion in terms of the plane waves of $\exp(\pm i k^\mu x_\mu)$. In a curved spacetime with the metric tensor of $g_{\mu\nu}$, Eq. (2) is no longer valid. However, one can expand an *arbitrary* field of $A^\mu(x)$ by means of its Fourier transform:

$$A^\mu(x) = \sum_k \sum_{s=0}^{3} \frac{1}{\sqrt{2\omega V}} e^\mu(\boldsymbol{k}, s)[f(\boldsymbol{k},s)\exp(-ik_\mu x^\mu) + f^\dagger(\boldsymbol{k},s)\exp(ik_\mu x^\mu)], \quad (11)$$

where $f(\boldsymbol{k},s)$ is an operator function in the momentum space (note that $A^\mu(x)$ is real). The Lorentz gauge condition of $\langle \Phi | \partial^\mu A_\mu | \Phi \rangle = 0$ expressed in an inertial coordinate system in flat spacetime now becomes

$$\langle \Phi | \nabla^\mu A_\mu | \Phi \rangle = \langle \Phi | (1/\sqrt{-g}) \partial^\mu (\sqrt{-g} A_\mu) | \Phi \rangle = 0, \quad (12)$$

where $g = \det(g_{\mu\nu})$, $\nabla^\mu$ denotes the covariant derivative. Eq. (12) implies that

$$\langle \Phi | \nabla^\mu A_\mu | \Phi \rangle = \langle \Phi | \partial^\mu A_\mu | \Phi \rangle + \langle \Phi | A_\mu \partial^\mu (\ln|g|) | \Phi \rangle = 0. \quad (13)$$

On the other hand, in the curved spacetime the electromagnetic field tensor satisfies

$$F_{\mu\nu} = \nabla_\mu A_\nu - \nabla_\nu A_\mu = \partial_\mu A_\nu - \partial_\nu A_\mu. \quad (14)$$

Thus $\boldsymbol{E} = -\nabla A^0 - \partial_t \boldsymbol{A}$ and $\boldsymbol{B} = \nabla \times \boldsymbol{A}$ are still valid. Similar to the process of obtaining

Eq. (5) from Eq. (2), from Eq. (11) one can obtain

$$\begin{cases} \boldsymbol{E} = \sum_{\boldsymbol{k},\lambda} \sqrt{\frac{\omega}{V}} [\frac{1}{\sqrt{1+\lambda^2}} \varepsilon(\boldsymbol{k},\lambda)][h(\boldsymbol{k},\lambda)\exp(-ik_\mu x^\mu) + h^\dagger(\boldsymbol{k},\lambda)\exp(ik_\mu x^\mu)] \\ \boldsymbol{B} = \sum_{\boldsymbol{k},\lambda} \sqrt{\frac{\omega}{V}} [\frac{-i}{\sqrt{1+\lambda^2}} \lambda \varepsilon(\boldsymbol{k},\lambda)][h(\boldsymbol{k},\lambda)\exp(-ik_\mu x^\mu) + h^\dagger(\boldsymbol{k},\lambda)\exp(ik_\mu x^\mu)] \end{cases}, \quad (15)$$

where $\lambda = \pm 1, 0$, and

$$\begin{cases} h(\boldsymbol{k},1) = if(\boldsymbol{k},1) \\ h(\boldsymbol{k},-1) = if(\boldsymbol{k},2) \\ h(\boldsymbol{k},0) = i[f(\boldsymbol{k},3) - f(\boldsymbol{k},0)]/\sqrt{2} \end{cases}. \quad (16)$$

For simplicity, let us assume that $g_{\mu\nu}$ is a diagonal matrix. In the present case, Eq. (10) becomes as [22]

$$\boldsymbol{J} = \int (\boldsymbol{E} \times \boldsymbol{B}) \sqrt{g_{11} g_{22} g_{33}} \, d^3 x. \quad (17)$$

To present a heuristic insight, let us take a weak field approximation as follows:

$$g_{\mu\nu} \approx \eta_{\mu\nu} + h_{\mu\nu}, \ h_{\mu\nu} \ll 1, \ h_{\mu\nu} \equiv 0 \text{ for } \mu,\nu \neq 0. \quad (18)$$

Assume that such a weak field approximation results in $f(\boldsymbol{k},s) \approx \hat{b}(\boldsymbol{k},s)$ (using Eqs. (6) and (16) one has $h(\boldsymbol{k},\lambda) \approx \hat{a}(\boldsymbol{k},\lambda)$), such that Eqs. (11), (15) and (16) can be replaced by Eqs. (2), (5) and (6), respectively. In particular, our weak field approximation given by Eq. (18) implies that $g_{11} = g_{22} = g_{33} = 1$, such that Eq. (17) is identical with Eq. (10). As a result, we can discuss the ZB of photons based on Eqs. (10), (7) and (13). In the stronger conditions of $\nabla^\mu A_\mu^{(+)} |\Phi\rangle = 0$, Eqs. (13) and (18) imply that

$$\partial^\mu A_\mu^{(+)} |\Phi\rangle + A_\mu^{(+)} \partial^\mu \ln(1+h_{00}) |\Phi\rangle = 0. \quad (19)$$

For the moment, Eq. (7) is no longer valid as long as the second term on the left-hand side of Eq. (19) does not vanish, it follows that the ZB contributions in Eq. (10) do not vanish provided that the gravitational interaction is taken into account.

As is well known, from the point of view of quantum field theory, gravitational forces come from the exchange of *virtual* gravitions, and a static gravitational field is full of *virtual* gravitions with the spin projections of $\pm 2$, $\pm 1$ and 0 (while the spin projections of a real gravition are $\pm 2$ only), where a virtual gravition can produce a virtual photon-photon pair (the anti-particle of a photon is the photon itself). For the moment, the ZB of photons can be interpreted as follows: assume that in a gravitational field there is a real photon (as the original transverse photon), around which a virtual photon pair consisting of a transverse and longitudinal (or scalar) photons (see the ZB terms in Eq. (10)), subsequently the longitudinal (or scalar) photon of the virtual pair annihilates with the original transverse photon, while the pair's transverse photon which is left over now replaces the original transverse photon, by such an exchange the ZB of the real transverse photon occurs.

## 4. Discussions and Conclusions

At the level of quantum field theory, the ZB of relativistic particles can be studied in a unified manner via the particles' momentum vectors (rather than via their position operators). Likewise, applying the volume-integrated Poynting vector, one can present a quantum field-theory investigation on the ZB of photons. The photonic ZB occurs only in the presence of virtual longitudinal and scalar photons. As a result, only the gravitational interaction can result in the ZB of photons. The ZB of a free relativistic particle has never been observed, but the behavior of such a particle has been simulated with a trapped ion, by putting it in an environment such that the non-relativistic Schrödinger equation for the ion has the same mathematical form as the Dirac equation (although the physical situation is different). Similarly, ZB might be studied by means of interdisciplinary studies [23-26].